\documentclass[journal]{IEEEtran}

\ifCLASSINFOpdf
\else
\fi

\usepackage{amsmath,amssymb,amsfonts}
\usepackage{graphicx}
\usepackage{textcomp}
\usepackage{wrapfig}
\usepackage{booktabs}
\usepackage{multirow}
\usepackage{siunitx}
\usepackage{bm}
\usepackage{tabularx}
\usepackage{array}
\usepackage{url}
\usepackage{hyperref}
\hypersetup{colorlinks=true, linkcolor=blue, citecolor=blue, urlcolor=blue}
\usepackage{xcolor}
\usepackage{makecell}
\usepackage[table]{xcolor}

\begin{document}

\title{Thermally Self-Compensated Differential Dual-Microring Sensor on Twin SiO$_2$ Membranes for Micronewton Force Probing}

\author{Phichai Youplao, Kittiwath Jeebkaew, and Sawatsakorn Chaiyasoonthorn~

\thanks{This work has been submitted to the IEEE for possible publication. Copyright may be transferred without notice, after which this version may no longer be accessible.}
}


\maketitle

\begin{abstract}
A silicon-photonic micronewton force sensor is proposed and numerically analyzed, in which two nominally matched all-pass microrings on one bus waveguide are placed on twin released SiO$_2$ membranes, and only one membrane is loaded by a probe. The differential resonance shift cancels the thermo-optic and thermomechanical drift common to both rings. The model couples a finite-difference mode solver, the anisotropic photoelastic tensor of silicon, transfer-matrix spectra, and a laminated-plate finite-element (FE) model including the ring stiffener and the compressive prestress of the buried oxide. The model agrees with an independent continuum FE model within 2.5\%, and a non-axisymmetric model including the bus waveguide predicts a 0.88\% lower sensitivity. A prestress-aware design rule (buckling factor ${\geq}3$) selects a 5-\si{\micro\meter} ring on a 17.5-\si{\micro\meter}-radius, 2-\si{\micro\meter}-thick membrane, giving a bus-inclusive sensitivity of $-1.50$~pm/\si{\micro\newton}, a loaded $Q$ of $9.8\times10^4$, nonlinearity below 0.05\% over 0--50~\si{\micro\newton}, a fitting-limited $3\sigma$ detection limit of 0.27~\si{\micro\newton} at 30-dB signal-to-noise ratio (0.34~\si{\micro\newton} at system level under the assumed 1-mK rms inter-membrane temperature fluctuation and optical-power conditions), and a residual thermal cross-sensitivity of $-0.13$~\si{\micro\newton}/K, 424 times lower than for a single ring. Monte Carlo analyses of fabrication, coupler, and resonance-separation tolerances and a comparison with a Vernier configuration are reported.

\end{abstract}

\begin{IEEEkeywords}
Differential sensing, force sensor, microring resonator, photoelastic effect, silicon photonics, temperature compensation.
\end{IEEEkeywords}

\IEEEpeerreviewmaketitle

\section{Introduction}
\label{sec:intro}
\IEEEPARstart{F}{orce} measurement at the micronewton level is required in cell manipulation, tissue palpation, and the characterization of microstructures \cite{Sun2002, Zang2023, Kumat2024}. Capacitive microelectromechanical (MEMS) sensors fabricated on silicon-on-insulator (SOI) wafers reach resolutions of 0.01~\si{\micro\newton} over ranges of several hundred micronewtons \cite{Sun2002}, but their electrical readout is susceptible to electromagnetic interference and requires one wired channel per element. Fiber Bragg grating (FBG) cantilevers resolve about 7~\si{\micro\newton} \cite{Zang2023}; fiber-tip Fabry--P\'erot probes printed by two-photon polymerization have reached a 54.9~nN detection limit \cite{Zou2021}; whispering-gallery-mode (WGM) microbubble and crystalline resonators have been demonstrated as force probes \cite{Liu2024, Wang2024}; and optical levers offer micronewton resolution with free-space optics \cite{Cheng2022}. These devices, however, are assembled individually and are not readily multiplexed on a chip.

Silicon microring resonators combine high quality factors, small footprint, and wafer-scale fabrication \cite{Bogaerts2012}. Their resonance wavelength responds to mechanical strain through the elongation of the optical path, the deformation of the waveguide cross section, and the photoelastic effect \cite{Westerveld2012, Westerveld2014}. When a ring is placed on a released membrane, the membrane converts a load into strain at the ring, as exploited in optical micromachined ultrasound sensors \cite{Leinders2015, Westerveld2021, Cao2026}. Two obstacles remain for quasi-static force sensing. First, the thermo-optic response of silicon \cite{Komma2012} produces a drift of about 80~pm/K, which for any realistic force sensitivity corresponds to tens of micronewtons per kelvin. Reference rings are routinely used to cancel this drift in refractive-index biosensors \cite{Iqbal2010}. Microring force sensing has recently been demonstrated with polymer-cladded silicon rings on a rigid substrate, reaching 66~fm/\si{\micro\newton} and a resolution of 12~\si{\micro\newton} in a two-dimensional array \cite{Safarloo2026}, and a reference ring outside the loaded region has been used to cancel temperature-induced resonance shifts of such a sensor in real time \cite{Safarloo2025}. On a released membrane, however, a reference ring cancels thermal drift only if it reproduces the thermomechanical strain and prestress state of the loaded membrane. Second, the buried oxide (BOX) of SOI wafers is under a compressive prestress of about $-300$~MPa \cite{Malhaire2023}, which can buckle released membranes and strongly modifies their stiffness.

This work addresses both obstacles with a differential dual-microring architecture on twin SiO$_2$ membranes, analyzed with a fully reproducible multiphysics model. The contributions are as follows.
\begin{enumerate}
\item A twin-membrane differential architecture in which the unloaded reference ring sits on its own, nominally identical released membrane, so that, unlike a reference on the rigid substrate \cite{Safarloo2025}, it tracks both the thermo-optic and the thermomechanical drift, reducing the thermal cross-sensitivity by more than two orders of magnitude.
\item A coupled model chaining a mode solver, a laminated-plate FE model with the ring stiffener and BOX prestress, the anisotropic photoelastic tensor, and transfer-matrix spectra, verified against independent axisymmetric continuum and non-axisymmetric plate FE models, the latter including the bus waveguide.
\item A prestress-aware design rule based on the linear buckling factor, together with an analysis showing that, under the adopted stability and laser-span constraints, Vernier magnification does not exceed the differential sensitivity.
\item A performance assessment covering detection limit, noise and self-heating, thermal drift, misalignment, nonlinearity, dynamics, fabrication, coupler, and resonance-separation Monte Carlo, and numerical convergence.
\end{enumerate}

\section{Sensor Architecture}
\label{sec:arch}
\begin{figure}[!t]
\centering
\includegraphics[width=3.4in]{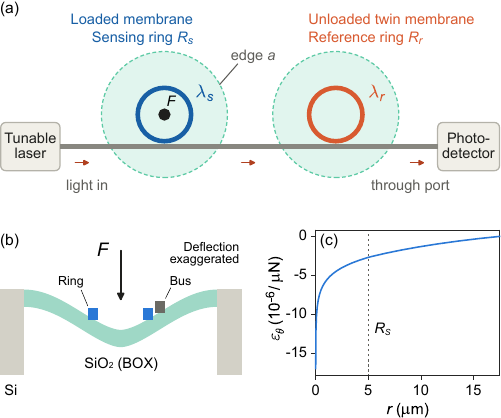}
\caption{(a) Layout and light path (not to scale): light from a tunable laser passes the sensing and reference microrings, each concentric with its own released membrane (dashed), along one bus waveguide and is detected at the through port. (b) Cross section of the loaded membrane through the ring center (deflection exaggerated; gap not to scale). (c) Simulated hoop strain at the ring centroid height per micronewton versus radius; the dashed line marks $R_s$.}
\label{fig:arch}
\end{figure}

Fig.~\ref{fig:arch}(a) depicts the sensor layout and the light path. Two all-pass microrings of radius $R_s=5.00$~\si{\micro\meter} (sensing) and $R_r=5.02$~\si{\micro\meter} (reference) are coupled to one bus waveguide. Each ring is concentric with its own circular membrane of radius $a=17.5$~\si{\micro\meter}, formed by the 2-\si{\micro\meter}-thick BOX after backside removal of the handle silicon \cite{Leinders2015}. The waveguides are 500~nm $\times$ 220~nm silicon strips with air above and SiO$_2$ below. Light from a tunable laser passes the sensing ring and then the reference ring and is detected at the through port. Each ring couples only to the bus, which crosses its membrane along a chord tangent to the ring. Near a resonance, light builds up in the ring; because the ring operates close to critical coupling, the transmitted and out-coupled fields interfere destructively and the power is dissipated by propagation loss, producing a transmission dip. One sweep therefore yields two dips, at $\lambda_s$ and $\lambda_r$. For equal mode order, the radius offset separates them by $\Delta\lambda\approx\lambda(\Delta R/R)(n_{\mathrm{eff}}/n_g)=3.64$~nm, so that they do not overlap over the force and temperature range. As illustrated in Fig.~\ref{fig:arch}(b), a rigid probe tip, such as a nanoindenter \cite{Safarloo2026}, pushes the center of the sensing membrane, so that the ring-side surface is compressed and $\lambda_s$ decreases. Fig.~\ref{fig:arch}(c) shows the resulting hoop strain at the ring-centroid height versus radius. The unloaded reference membrane experiences the same temperature, prestress, and thermomechanical strain.

\section{Modeling Framework}
\label{sec:model}

\subsection{Optical Model}
\begin{figure*}[!t]
\centering
\includegraphics[width=7.0in]{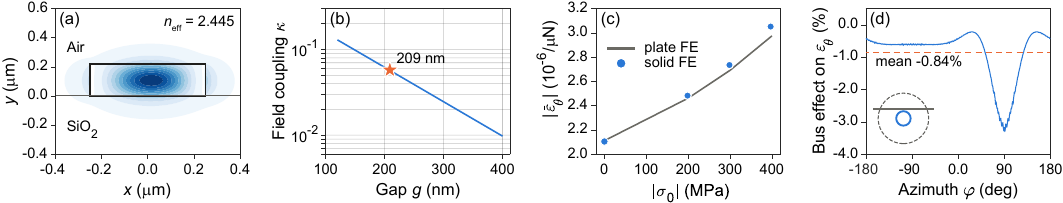}
\caption{(a) Quasi-TE mode intensity of the bent waveguide ($R=5$~\si{\micro\meter}). (b) Bus--ring field coupling versus gap; the star marks the design. (c) Ring hoop strain per micronewton versus BOX prestress from the laminated-plate model and the axisymmetric continuum FE model. (d) Change of the ring hoop strain caused by the bus waveguide versus azimuth (bus tangent at $\varphi=90^\circ$; mean $-0.84\%$) from the non-axisymmetric plate FE model (0.175-\si{\micro\meter} grid, running mean over one element); inset: geometry.}
\label{fig:model}
\end{figure*}

The quasi-TE mode of the bent waveguide is computed with a semi-vectorial finite-difference solver \cite{Stern1988} on a 10-nm grid, with the bend represented by the conformal transformation $n^2(x)\rightarrow n^2(x)(1+x/R)^2$ \cite{Heiblum1975}. At $R=5$~\si{\micro\meter} and $\lambda=1.55$~\si{\micro\meter}, the guided quasi-TE mode has $n_{\mathrm{eff}}=2.445$ and $n_g=n_{\mathrm{eff}}-\lambda\,dn_{\mathrm{eff}}/d\lambda=4.153$, and 71.7\% of the mode power lies in silicon; the mode intensity is shown in Fig.~\ref{fig:model}(a). Refining the grid to 5~nm changes $n_{\mathrm{eff}}$ by less than 0.3\% and the force sensitivity $S_F$ by less than 1\%. The solver returns $n_{\mathrm{eff}}$ about 2\% above full-vectorial values, which changes $S_F$ by 3.2\%; this is retained as a model uncertainty.

The through-port transmission of the two cascaded all-pass rings is
\begin{equation}
T(\lambda)=\left|\tau_s\tau_r\right|^2,\quad
\tau_i=\frac{t_i-a_ie^{-j\phi_i}}{1-t_ia_ie^{-j\phi_i}},\quad
\phi_i=\frac{2\pi n_{\mathrm{eff}}(\lambda)L_i}{\lambda},
\label{eq:tmm}
\end{equation}
where $t_i=\sqrt{1-\kappa_i^2}$ with the gap-dependent field coupling $\kappa_i$ shown in Fig.~\ref{fig:model}(b) (Section~\ref{sec:coupler}), $a_i$ is the round-trip amplitude transmission for a propagation loss of 3~dB/cm \cite{Bogaerts2012}, and $L_i=2\pi R_i$ \cite{Yariv2000}. A perturbation of a ring enters (\ref{eq:tmm}) as an effective-index change, so that the shift scales with the resonance wavelength. For a ring with azimuthally averaged hoop strain $\bar\varepsilon_\theta$ and accompanying radial and vertical strains $\bar\varepsilon_r$ and $\bar\varepsilon_z$, the resonance shift is \cite{Westerveld2012}
\begin{equation}
\frac{\Delta\lambda}{\lambda}=\frac{1}{n_g}\Big[n_{\mathrm{eff}}\bar\varepsilon_\theta
+\frac{\partial n_{\mathrm{eff}}}{\partial w}w\bar\varepsilon_r
+\frac{\partial n_{\mathrm{eff}}}{\partial h}h\bar\varepsilon_z
+\frac{\partial n_{\mathrm{eff}}}{\partial n_{\mathrm{Si}}}\Delta n_{\mathrm{Si}}\Big],
\label{eq:shift}
\end{equation}
with the terms describing path elongation, cross-section change, and the photoelastic index change, and $\partial n_{\mathrm{eff}}/\partial n_{\mathrm{Si}}=1.103$, $\partial n_{\mathrm{eff}}/\partial w=1.52$~\si{\per\micro\meter}, $\partial n_{\mathrm{eff}}/\partial h=3.62$~\si{\per\micro\meter} from the mode solver. Because $n_g>n_{\mathrm{eff}}$, the geometric term is scaled by $n_{\mathrm{eff}}/n_g=0.59$.

\subsection{Elasto-Optic Model}
\label{sec:elas_model}
The ring is uniaxially stressed in the hoop direction, which rotates through the crystal around the ring on a (001) wafer. The strain tensor $\bm{\varepsilon}(\varphi)$ is therefore obtained from the cubic compliance of silicon \cite{Hopcroft2010} at each azimuth $\varphi$, and the index change seen by the radially polarized TE field is
\begin{equation}
\Delta n_{\mathrm{Si}}=-\frac{n_{\mathrm{Si}}^3}{2}\left\langle\hat{\mathbf r}^{\mathsf T}\left(\mathbf p:\bm{\varepsilon}(\varphi)\right)\hat{\mathbf r}\right\rangle_\varphi,
\label{eq:pe}
\end{equation}
with $p_{11}=-0.101$, $p_{12}=0.0094$, and $p_{44}=-0.051$ \cite{Biegelsen1974}. Averaging gives $\bar\varepsilon_r=-0.185\,\bar\varepsilon_\theta$, $\bar\varepsilon_z=-0.315\,\bar\varepsilon_\theta$, $\Delta n_{\mathrm{Si}}=-0.487\,\bar\varepsilon_\theta$, and an effective hoop modulus of 147~GPa. Under compression ($\bar\varepsilon_\theta<0$) the photoelastic and cross-section terms are positive and oppose the dominant geometric term; the net shift is 0.62 times the geometric contribution.

\subsection{Membrane Mechanics}
The membrane is modeled as an axisymmetric Kirchhoff laminate with radial displacement $u(r)$ and deflection $w(r)$, discretized with Hermite elements for $w$ and linear elements for $u$ \cite{Timoshenko1959, Reddy2007}. The total potential energy is
\begin{multline}
\Pi=\pi\!\int_0^a\!\left(\bm\epsilon_0^{\mathsf T}\mathbf A\bm\epsilon_0+2\bm\epsilon_0^{\mathsf T}\mathbf B\bm\kappa+\bm\kappa^{\mathsf T}\mathbf D\bm\kappa+N_0w'^2\right)r\,dr\\
+\pi R E_rA_r\left(\varepsilon_{\mathrm{ring}}-\alpha_{\mathrm{Si}}\Delta T\right)^2+F\,w(0),
\label{eq:energy}
\end{multline}
with $\bm\epsilon_0=[u',\,u/r]^{\mathsf T}$, $\bm\kappa=-[w'',\,w'/r]^{\mathsf T}$, the laminate stiffness matrices $\mathbf A$, $\mathbf B$, $\mathbf D$ (SiO$_2$: $E=73$~GPa, $\nu=0.17$), and the prestress resultant $N_0=\sigma_0t$ with $\sigma_0=-300$~MPa \cite{Malhaire2023}. The deflection $w$ is positive upward (toward the ring), and $F>0$ acts downward. The silicon ring is a discrete hoop stiffener of axial rigidity $E_rA_r$ with eigenstrain $\alpha_{\mathrm{Si}}\Delta T$ and centroid strain
\begin{equation}
\varepsilon_{\mathrm{ring}}=\frac{u(R)-z_c\,w'(R)}{R},
\label{eq:ring}
\end{equation}
where $z_c=t+h/2$. All through-thickness integrals ($\mathbf A$, $\mathbf B$, $\mathbf D$, thermal resultants) use $z$ measured from the bottom of the BOX, which is thus the reference plane of $u(r)$; hence $\mathbf B\neq\mathbf 0$ even for the single layer, and $z_c$ is measured from the same plane. A mid-plane formulation ($\mathbf B=\mathbf 0$, $z_c=t/2+h/2$) changes the ring strain by $1\times10^{-6}$ (relative), and 400 elements agree with 800 elements to $3\times10^{-6}$. Stability is quantified by the smallest $\lambda_{\mathrm{cr}}$ satisfying
\begin{equation}
\det\left[\mathbf K_e+\lambda_{\mathrm{cr}}\mathbf K_g(N_0)\right]=0,
\label{eq:buckle}
\end{equation}
where $\mathbf K_e$ and $\mathbf K_g$ are the elastic and geometric stiffness matrices \cite{Timoshenko1961}. Temperature enters through the thermal resultants of the layers, the ring eigenstrain, and the substrate-imposed edge displacement $u(a)=\alpha_{\mathrm{Si}}a\Delta T$.

\subsection{Independent Verification}
Two independent models verify the plate model. First, an axisymmetric continuum FE model of eight-node quadrilaterals was built, in which the ring cross section is meshed explicitly, the load acts over a 0.25-\si{\micro\meter} radius, and the prestress enters through the initial-stress stiffness \cite{Zienkiewicz2000}. As shown in Fig.~\ref{fig:model}(c), its ring strain agrees with the plate model within $-0.46\%$ to $+2.46\%$ for prestress between 0 and $-400$~MPa, and within 2.4\% for other geometries. Second, because the straight bus crosses the membrane, axisymmetry was tested with a non-axisymmetric laminated-plate FE model on a Cartesian grid (Bogner--Fox--Schmit elements for $w$, bilinear elements for the in-plane displacements) in which the ring (effective hoop modulus 147~GPa, Section~\ref{sec:elas_model}) and the bus ($E=169$~GPa along $\langle110\rangle$) are bonded axial line stiffeners at height $z_c$. The ring is sampled at 2048 azimuthal points, where the axial strain $\varepsilon_\theta(\varphi)=\hat{\mathbf t}^{\mathsf T}\nabla\mathbf u\,\hat{\mathbf t}-z_c\,\hat{\mathbf t}^{\mathsf T}\nabla\nabla w\,\hat{\mathbf t}$ is interpolated from the element fields, and the optical shift uses $\bar\varepsilon_\theta=(2\pi)^{-1}\oint\varepsilon_\theta\,d\varphi$, consistent with (\ref{eq:shift}); the bus continues into the clamped region and is thus anchored at the membrane edge. Without the bus, this model reproduces the axisymmetric ring strain within 0.01\% at a 0.175-\si{\micro\meter} grid. As shown in Fig.~\ref{fig:model}(d), the bus reduces the ring strain by up to about 3\% near the tangent point and by 0.84\% on average, a value converged within 0.05\% between 0.35- and 0.175-\si{\micro\meter} grids. As the ring stress stays uniaxial, the relations of Section~\ref{sec:elas_model} hold locally, and since (\ref{eq:shift}) is linear in the local perturbation, weighting the strain change with the orientation-dependent optical response (1.20--1.92 per unit hoop strain around the ring) gives a sensitivity change of $-0.88\%$ ($-0.81\%$ for $\langle100\rangle$). The bus-inclusive sensitivity, $S_F=-1.50$~pm/\si{\micro\newton}, is taken as that of the proposed device; the buckling factor changes from 3.03 to 3.05. Because the bus loading under temperature is common to both membranes, the 2-D model was used for the force response only. Force-equivalent values in Section~\ref{sec:results} are computed with the axisymmetric $S_F=-1.514$~pm/\si{\micro\newton} and are 0.9\% larger for the bus-inclusive device.

\subsection{Differential Readout and Detection Limit}
The differential output is
\begin{equation}
\Delta\lambda_D=\Delta\lambda_s-\Delta\lambda_r=S_F F+\left(\beta_s-\beta_r\right)\Delta T,
\label{eq:diff}
\end{equation}
where $\beta_{s,r}$ are the temperature coefficients of the two resonances. Laser wavelength jitter is common to both resonances, which are acquired in the same sweep, and cancels. The detection limit is
\begin{equation}
F_{\min}=3\sigma_D/|S_F|,
\label{eq:lod}
\end{equation}
where $\sigma_D$ is the standard deviation of the fitted differential shift, obtained by fitting Lorentzian line shapes to spectra with white amplitude noise of standard deviation $10^{-\mathrm{SNR}/20}$ per sample (1-pm sampling, 300 realizations) \cite{White2008}. The reference interrogation is a 10-nm/s sweep (10~kS/s), so that a line width is crossed in 1.6~ms and one differential reading over a 5-nm window takes 0.5~s; static offsets are removed by taring at $F=0$. Shot and detector noise (1~pW/$\sqrt{\mathrm{Hz}}$) allow over 80~dB per sample even at 1~\si{\micro\watt} received, so the assumed 30--40~dB reflect practical coupling and fringe noise.

\section{Design Optimization}
\label{sec:design}
\subsection{Ring and Membrane Geometry}
\begin{figure}[!t]
\centering
\includegraphics[width=3.4in]{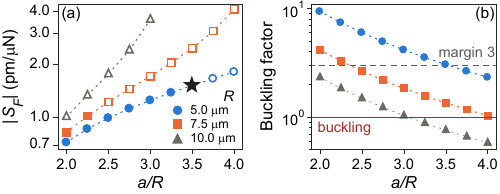}
\caption{Design map at $t=2$~\si{\micro\meter}. (a) Force sensitivity versus $a/R$ (prestress $-300$~MPa); filled markers satisfy the stability margin, open markers do not, and the star marks the selected design. (b) Linear buckling factor at $-400$~MPa prestress.}
\label{fig:design}
\end{figure}

Without prestress, the ring strain scales as $\ln(a/R)/t^2$ and is independent of $R$ for fixed $a/R$. The prestress breaks this invariance: larger membranes approach buckling, where the sensitivity diverges and depends strongly on the uncertain prestress. Fig.~\ref{fig:design}(a) and (b) show the force sensitivity and the linear buckling factor at a worst-case prestress of $-400$~MPa, respectively, versus $a/R$ for $R=5$, $7.5$, and $10$~\si{\micro\meter}. Designs were accepted if $\lambda_{\mathrm{cr}}\geq3$ at this prestress and if $S_F$ varied by no more than 25\% between $-200$ and $-400$~MPa. The most sensitive admissible point is $R=5$~\si{\micro\meter}, $a/R=3.5$ ($\lambda_{\mathrm{cr}}=3.05$, sensitivity spread 19\%); larger rings, which would offer a smaller free spectral range (FSR), are excluded. The parameters are listed in Table~\ref{tab:params}.

\begin{table}[!t]
\renewcommand{\arraystretch}{1.12}
\caption{Design Parameters}
\footnotesize
\setlength{\tabcolsep}{3pt}
\label{tab:params}
\centering
\begin{tabularx}{\columnwidth}{@{} p{5cm} l @{}}
\toprule
Parameter & Value \\
\midrule
Waveguide (Si strip, air top, SiO$_2$ bottom) & 500~nm $\times$ 220~nm \\
Ring radii $R_s$ / $R_r$ & 5.00 / 5.02~\si{\micro\meter} \\
Membrane radius $a$ / thickness $t$ (BOX) & 17.5 / 2.0~\si{\micro\meter} \\
BOX prestress $\sigma_0$ (nominal / range) & $-300$ / $-200$ to $-400$~MPa \\
Bus--ring gap / field coupling $\kappa$ & 209~nm / 0.057 \\
Propagation loss & 3~dB/cm \\
$n_{\mathrm{eff}}$ / $n_g$ at $R_s$ & 2.445 / 4.153 \\
SiO$_2$: $E$, $\nu$, $\alpha$ & 73~GPa, 0.17, 0.5~ppm/K \\
Si: $p_{11}$, $p_{12}$, $p_{44}$ \cite{Biegelsen1974} & $-0.101$, 0.0094, $-0.051$ \\
Si: $dn/dT$ \cite{Komma2012}, $\alpha$ & $1.86\times10^{-4}$~K$^{-1}$, 2.6~ppm/K \\
Si / SiO$_2$: $\rho$ (kg/m$^3$) & 2330 / 2200 \\
Si / SiO$_2$: $c_p$ (J\,kg$^{-1}$K$^{-1}$) & 700 / 730 \\
\bottomrule
\end{tabularx}
\end{table}

\subsection{Coupler Design}
\label{sec:coupler}
The bus--ring coupling is computed from the even--odd supermode splitting of two parallel waveguides, $\kappa_s(g)=\pi\Delta n_{\mathrm{sm}}(g)/\lambda\approx\kappa_0e^{-\gamma g}$, integrated along the parabolic gap profile $g(x)\approx g+x^2/(2R)$ \cite{Little1997}:
\begin{equation}
\kappa=\sin\left[\kappa_0e^{-\gamma g}\sqrt{2\pi R/\gamma}\right],
\label{eq:coupler}
\end{equation}
with $\kappa_0=0.214$~\si{\per\micro\meter} and $\gamma=9.25$~\si{\per\micro\meter}. A slightly over-coupled $\kappa=0.057$ ($\kappa^2=1.5(1-a_i^2)$), which balances extinction and line width, requires $g=209$~nm, as marked in Fig.~\ref{fig:model}(b); a $\pm10$-nm gap error changes $\kappa$ to 0.063--0.052, and gap variations are included in the Monte Carlo of Section~\ref{sec:fab}.

\subsection{Comparison With a Vernier Configuration}
A Vernier cascade magnifies the shift by $M=\mathrm{FSR}_r/|\mathrm{FSR}_r-\mathrm{FSR}_s|$ \cite{Claes2010}, but its envelope period $\mathrm{FSR}_V\approx M\,\mathrm{FSR}_s$ must fit within the laser span, here taken as 35~nm. For $M=8$ this requires $R\geq20$~\si{\micro\meter} and $a\geq30$~\si{\micro\meter}, which gives $\lambda_{\mathrm{cr}}=0.48$--1.03 at $-400$~MPa on a 2-\si{\micro\meter} BOX and only 1.1--2.3 on a 3-\si{\micro\meter} BOX. Satisfying both constraints requires $M=5.5$, $R=15$~\si{\micro\meter} ($\mathrm{FSR}_V=33.9$~nm), $a/R=1.5$, and a 3-\si{\micro\meter} BOX give $\lambda_{\mathrm{cr}}=4.1$ but a single-ring sensitivity of $-0.215$~pm/\si{\micro\newton}, so that even the ideal envelope shift $MS_{\mathrm{ring}}=-1.18$~pm/\si{\micro\newton} remains below the $-1.51$~pm/\si{\micro\newton} of the axisymmetric differential design, while the envelope, broadened by $M$, is fitted less precisely \cite{Claes2010}; thermal behavior does not discriminate, since matched Vernier membranes can also be compensated. The differential configuration is therefore preferred under the constraints adopted here, which is a result for this platform rather than a general statement on Vernier sensors.

\section{Results and Discussion}
\label{sec:results}

\subsection{Force Response}
\begin{figure}[!t]
\centering
\includegraphics[width=3.4in]{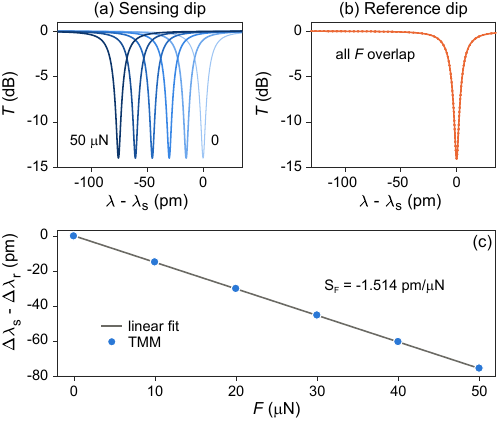}
\caption{Noise-free TMM response. (a) Sensing dip for $F=0$--50~\si{\micro\newton} in steps of 10~\si{\micro\newton}. (b) Reference dip; all six curves coincide. (c) Differential shift versus force with a linear fit (axisymmetric baseline, $-1.514$~pm/\si{\micro\newton}).}
\label{fig:force}
\end{figure}

Fig.~\ref{fig:force}(a) shows the noise-free spectra for $F=0$--50~\si{\micro\newton}. The sensing dip (1541.93~nm) shifts by $-75.7$~pm at 50~\si{\micro\newton} (4.8 line widths), whereas Fig.~\ref{fig:force}(b) shows that the reference dip (1545.57~nm) is unchanged within the 0.1-pm sampling. The line shape is preserved: at 50~\si{\micro\newton} the center deflection is approximately of 7.7~nm, the bus deflects with the ring on the same membrane (relative vertical offset below 0.5~nm), and the gap change of 0.03~nm alters $\kappa$ by less than 0.03\%. The loaded $Q$ is $9.8\times10^4$ (line width 15.7~pm), the extinction 14.0~dB, and the FSR 18.4~nm. The linear fit in Fig.~\ref{fig:force}(c) gives an axisymmetric differential sensitivity of $S_F=-1.514$~pm/\si{\micro\newton}, within 0.5\% of (\ref{eq:shift}) at 1550~nm owing to the scaling $\lambda_s/1550$~nm. The contributions are $-2.46$ (path elongation), $+0.39$ (cross section), and $+0.54$~pm/\si{\micro\newton} (photoelastic).

\subsection{Detection Limit and Noise Budget}
\label{sec:limit}
\begin{figure}[!t]
\centering
\includegraphics[width=3.4in]{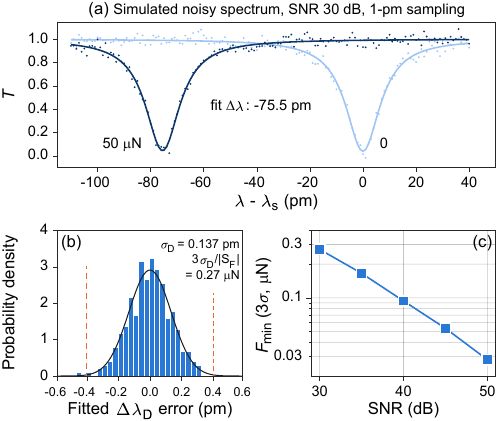}
\caption{Noise and detection limit. (a) One simulated noisy through-port spectrum (SNR 30~dB, 1-pm sampling) at $F=0$ and 50~\si{\micro\newton} with Lorentzian fits. (b) Distribution of the fitted differential-shift error over 300 Monte Carlo realizations at 30~dB with a Gaussian of the same standard deviation; dashed lines mark $\pm3\sigma_D$. (c) $3\sigma$ detection limit versus SNR.}
\label{fig:noise}
\end{figure}

Fig.~\ref{fig:noise}(a) shows one realization at 30~dB SNR, in which the fits recover $-75.5$~pm (model: $-75.7$~pm). As shown in Fig.~\ref{fig:noise}(b), the fitted differential shift over 300 realizations is Gaussian with $\sigma_D=0.137$~pm, giving $F_{\min}=0.27$~\si{\micro\newton}; at 40~dB, $\sigma_D=0.047$~pm and $F_{\min}=0.094$~\si{\micro\newton}, and Fig.~\ref{fig:noise}(c) shows the dependence on SNR. The Cram\'er--Rao bound of the Lorentzian fit reproduces these values within 8\%.

\begin{figure}[!t]
\centering
\includegraphics[width=3.4in]{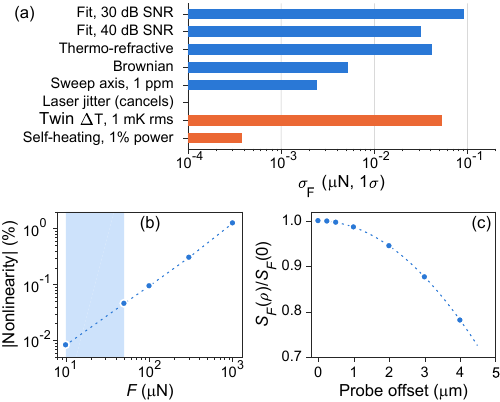}
\caption{(a) 1$\sigma$ force-equivalent noise budget of Table~\ref{tab:noise}; orange bars are fluctuations of operating conditions (inter-membrane temperature, optical power) rather than intrinsic noise. (b) Deviation of the von K\'arm\'an solution from the linear response (target range shaded). (c) Relative sensitivity versus lateral probe offset from Maxwell--Betti reciprocity.}
\label{fig:robust}
\end{figure}

The force-equivalent noise terms are listed in Table~\ref{tab:noise} and compared in Fig.~\ref{fig:robust}(a). The thermo-refractive noise is evaluated separately for the silicon core and the oxide, each with its mode-weighted thermal volume $V_i=L(\int_i I\,dA)^2/\int_i I^2\,dA$ and temperature variance $k_BT^2/(\rho_ic_iV_i)$ \cite{Gorodetsky2004}; silicon dominates ($V=2.3$~\si{\micro\meter\cubed}; rms temperature fluctuation of this mode-weighted volume 0.58~mK), giving 0.041~\si{\micro\newton} for two uncorrelated rings, whereas the oxide contributes only $1.5\times10^{-4}$~\si{\micro\newton}. The Brownian force noise of the membrane, $\sqrt{k_BTk_{\mathrm{eff}}}$ with $k_{\mathrm{eff}}=6.46$~kN/m, is 0.005~\si{\micro\newton}. Both are full-band values; their correlation times (about 0.5~ns in the core and 1~\si{\micro\second} in the oxide) and the 15.5-MHz membrane resonance are far faster than the 1.6-ms line-crossing time, so they are conservative bounds. A 1-ppm wavelength-axis nonlinearity over the dip separation contributes 0.002~\si{\micro\newton}. Static differences between the membranes, such as a constant temperature offset or unequal self-heating, appear as offsets removed by taring; only their fluctuations during a reading are treated as noise (self-heating is quantified in Section~\ref{sec:fab}). An rms fluctuation of 1~mK in the inter-membrane temperature difference averaged over one reading, assumed as an operating requirement rather than a demonstrated value, enters with the single-ring coefficient of 53~\si{\micro\newton}/K and gives 0.053~\si{\micro\newton}, the dominant practical term; a static gradient of 0.1~K/mm across a 40-\si{\micro\meter} membrane spacing would produce a 0.21-\si{\micro\newton} offset. The membranes should therefore be adjacent, and the probe thermalized before readout.

\begin{table}[!t]
\renewcommand{\arraystretch}{1.12}
\caption{Force-Equivalent Noise and Detection Limits}
\label{tab:noise}
\centering
\begin{tabularx}{\columnwidth}{@{} X c c @{}}
\toprule
Contribution (1$\sigma$, \si{\micro\newton}) & 30~dB & 40~dB \\
\midrule
Lorentzian fitting (Monte Carlo) & 0.090 & 0.031 \\
Thermo-refractive, Si + SiO$_2$, two rings$^{a}$ & \multicolumn{2}{c}{0.041} \\
Membrane Brownian$^{a}$ & \multicolumn{2}{c}{0.005} \\
Sweep-axis nonlinearity, 1~ppm & \multicolumn{2}{c}{0.002} \\
Inter-membrane $\Delta T$ fluctuation, 1~mK rms & \multicolumn{2}{c}{0.053} \\
Self-heating, 1\% power fluctuation & \multicolumn{2}{c}{0.0004} \\
\midrule
$F_{\min}$ (3$\sigma$), fitting-limited & 0.27 & 0.094 \\
$F_{\min}$ (3$\sigma$), intrinsic (first four rows) & 0.30 & 0.16 \\
$F_{\min}$ (3$\sigma$), system$^{b}$ (all rows) & 0.34 & 0.22 \\
\bottomrule
\multicolumn{3}{@{}p{\columnwidth}@{}}{\footnotesize Axisymmetric $S_F$. Conditions: SNR per 1-pm sample, 10-nm/s sweep, static offsets tared at $F=0$. Root-sum-square combination. $^{a}$Full-band value (upper bound). $^{b}$Conditional on the assumed operating conditions: 1-mK rms inter-membrane temperature fluctuation, absorbed power ${\leq}1$~\si{\micro\watt} per ring, 1\% absorption imbalance, 1\% power stability.}\\
\end{tabularx}
\end{table}

\subsection{Nonlinearity, Misalignment, and Dynamics}
As the linear model is linear by construction, linearity was assessed with a geometrically nonlinear (von K\'arm\'an) solution of (\ref{eq:energy}) by Newton iteration; as shown in Fig.~\ref{fig:robust}(b), it deviates from the linear response by $-0.008\%$, $-0.046\%$, and $-0.094\%$ at 10, 50, and 100~\si{\micro\newton}, and by $-1.3\%$ only at 1~mN; the linear range thus exceeds $100~\si{\micro\newton}$, more than 350 times the fitting-limited $F_{\min}$.

The response to an off-center load follows from Maxwell--Betti reciprocity without a three-dimensional model: the ring strain produced by a point load at radius $\rho$ equals the deflection at $\rho$ of the axisymmetric field $\mathbf w_c$ generated by the load vector conjugate to (\ref{eq:ring}),
\begin{equation}
\frac{S_F(\rho)}{S_F(0)}=\frac{w_c(\rho)}{w_c(0)},\qquad \mathbf K\mathbf w_c=\partial\varepsilon_{\mathrm{ring}}/\partial\mathbf q.
\label{eq:recip}
\end{equation}
As shown in Fig.~\ref{fig:robust}(c), the sensitivity decreases by 0.35\%, 1.4\%, and 5.5\% for offsets of 0.5, 1, and 2~\si{\micro\meter}, and by 0.7\% for a 1-\si{\micro\meter}-radius tip, so a 1-\si{\micro\meter} alignment tolerance suffices. The fundamental mechanical resonance of the prestressed membrane with the ring mass is 15.5~MHz (17.7~MHz without prestress); the measurement update rate, about two differential readings per second, is set by the laser sweep, not by the mechanics.

\subsection{Thermal Cross-Sensitivity}
Fig.~\ref{fig:thermal}(a) shows that, between 290 and 310~K, the single sensing resonance drifts by 80.3~pm/K, equivalent to 53.0~\si{\micro\newton}/K, including the thermo-optic effect and the substrate-imposed thermal strain of the ring ($2.6\times10^{-6}$~K$^{-1}$). In contrast, the differential output in Fig.~\ref{fig:thermal}(b) drifts by only $-0.19$~pm/K ($-0.13$~\si{\micro\newton}/K), a rejection factor of 424. The residual is deterministic, arising from the 0.24\% wavelength difference between the resonances, and can be removed by calibration. The thermal-stress change of the BOX gives a temperature coefficient of $S_F$ of $-175$~ppm/K, i.e., $\pm0.17\%$ over $\pm10$~K.

\subsection{Fabrication Tolerances}
\label{sec:fab}
\begin{figure}[!t]
\centering
\includegraphics[width=3.4in]{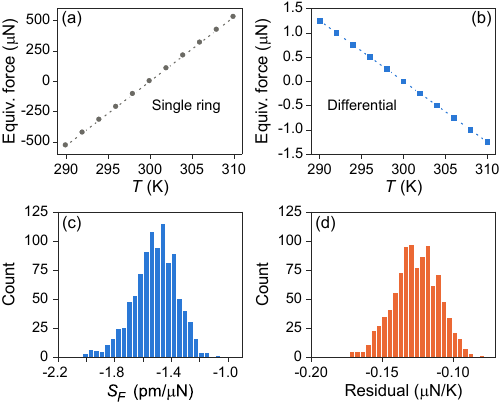}
\caption{Equivalent force versus temperature for (a) the single sensing ring and (b) the differential output. Monte Carlo distributions ($N=1000$) of (c) $S_F$ and (d) the residual differential thermal coefficient.}
\label{fig:thermal}
\end{figure}

\begin{figure}[!t]
\centering
\includegraphics[width=3.4in]{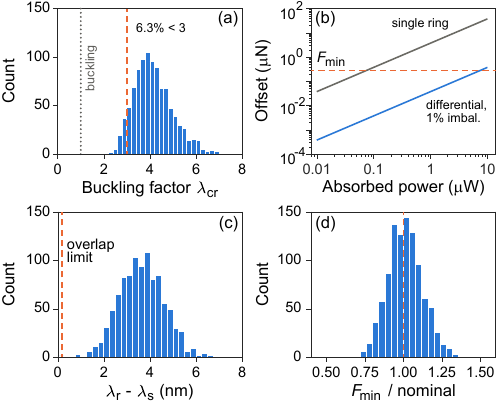}
\caption{Monte Carlo and operating-point analysis. (a) Buckling factor (smaller of the two membranes) of the $N=1000$ devices at their sampled prestress; dotted line: buckling, dashed line: design margin. (b) Static self-heating offset versus absorbed power for a single ring and for the differential output with 1\% absorption imbalance (removed by taring); dashed line: $F_{\min}$ at 30~dB. (c) Resonance separation; dashed line: overlap limit of 12 line widths. (d) Cram\'er--Rao-based $F_{\min}$ at 30~dB relative to the nominal design.}
\label{fig:stab}
\end{figure}

A Monte Carlo analysis ($N=1000$, independent Gaussian variables) applied global variations common to both rings (standard deviations: width 5~nm, thickness 2~nm, prestress 50~MPa, BOX thickness 2\%, membrane radius 1~\si{\micro\meter}) and local mismatches between the rings (width 1~nm, thickness 0.3~nm, membrane radius 0.2~\si{\micro\meter}, prestress 5~MPa). As shown in Fig.~\ref{fig:thermal}(c) and (d), the sensitivity is $S_F=-1.52\pm0.16$~pm/\si{\micro\newton} (10\% spread), which requires a one-point force calibration of each device. The residual thermal coefficient is $-0.126\pm0.016$~\si{\micro\newton}/K, with 95\% of the devices below 0.154~\si{\micro\newton}/K in magnitude. As shown in Fig.~\ref{fig:stab}(a), the buckling factor, evaluated for both membranes in each device at its prestress, has a median of 4.07 and a minimum of 2.38; no device buckles, but 6.3\% fall below the design margin of 3, giving a predicted acceptance yield of 93.7\% if the margin is enforced.

\begin{table*}[!t]
\renewcommand{\arraystretch}{1.12}
\caption{Comparison With Reported Micronewton Force Sensors}
\label{tab:compare}
\centering
\footnotesize
\setlength{\tabcolsep}{5pt}
\begin{tabularx}{\textwidth}{@{}l X c c c c c@{}}
\toprule
Ref. & Principle & Range & Sensitivity & Resolution/LOD & EMI immune & On-chip multiplex. \\
\midrule
\cite{Sun2002} & MEMS capacitive (comb drive) & ${\leq}490$~\si{\micro\newton} & --- & 0.01~\si{\micro\newton} & No & Electrical \\
\cite{Zang2023} & FBG cantilever & --- & --- & 7~\si{\micro\newton} & Yes & No \\
\cite{Zou2021} & Fiber-tip polymer Fabry--P\'erot & ${\sim}2.9$~mN & 1.51~nm/\si{\micro\newton} & 54.9~nN & Yes & No \\
\cite{Wang2024} & CaF$_2$ WGM resonator & --- & --- & 8.49~\si{\micro\newton} & Yes & No \\
\cite{Safarloo2026} & Polymer-cladded Si microring array (rigid substrate) & --- & 66~fm/\si{\micro\newton} & 12~\si{\micro\newton} & Yes & Yes \\
This work & Differential Si microrings on twin membranes & 0--50~\si{\micro\newton}$^{a}$ & $-1.50$~pm/\si{\micro\newton}$^{b}$ & 0.27~\si{\micro\newton}$^{c}$ & \,~Expected$^{d}$ & \,~Expected$^{d}$ \\
\bottomrule
\multicolumn{7}{@{}p{\textwidth}@{}}{\footnotesize $^{a}$Simulated (this work); linear within 0.1\% up to 100~\si{\micro\newton}. $^{b}$Bus-inclusive; the axisymmetric model gives $-1.514$~pm/\si{\micro\newton}. $^{c}$Fitting-limited, 3$\sigma$ at 30~dB SNR; 0.34~\si{\micro\newton} at system level under the assumed operating conditions of Table~\ref{tab:noise}; residual thermal cross-sensitivity $-0.13$~\si{\micro\newton}/K. $^{d}$EMI immunity is expected from the all-optical readout, whereas on-chip wavelength-domain multiplexing is feasible in principle; neither has been experimentally demonstrated. ---: not stated in the cited work.}\\
\end{tabularx}
\end{table*}

Self-heating follows from the radial conduction resistance of the membrane annulus, $R_{\mathrm{th}}=\ln(a/R)/(2\pi k_{\mathrm{ox}}t)=7.1\times10^4$~K/W ($k_{\mathrm{ox}}=1.4$~W\,m$^{-1}$K$^{-1}$, time constant 0.22~ms): as shown in Fig.~\ref{fig:stab}(b), one microwatt of absorbed power shifts a single resonance by 5.7~pm (3.8~\si{\micro\newton}). The differential output cancels this to the extent that both rings absorb equal power. A 1\% absorption imbalance, conservative since a 1-nm local gap mismatch changes the dissipated fraction by 0.36\%, leaves a tared offset of 0.036~\si{\micro\newton} per absorbed microwatt, and a 1\% power fluctuation at 1~\si{\micro\watt} contributes $4\times10^{-4}$~\si{\micro\newton}. An absorbed power ${\leq}1$~\si{\micro\watt} per ring is a design requirement, since the absorbed fraction of the loss is unknown before fabrication; it also keeps the thermal detuning at 0.35 line widths, avoiding bistability.

The coupler gap was added to the same width and thickness samples as an independent Gaussian variable (global 5~nm, local 1~nm). The field coupling is $0.057\pm0.003$, the line width $15.8\pm0.9$~pm, the extinction $13.9\pm1.8$~dB, and the loaded $Q$ $(9.8\pm0.5)\times10^4$. Fig.~\ref{fig:stab}(c) shows the resonance separation $\lambda_r-\lambda_s$, which is $3.65\pm0.96$~nm and dominated by the local width mismatch; its minimum over all devices, 0.84~nm, exceeds the 0.19-nm clearance (12 line widths) needed for non-overlapping fit windows, so that no device shows overlapping dips over the force and temperature range. Fig.~\ref{fig:stab}(d) shows that the Cram\'er--Rao-based $F_{\min}$, evaluated with the $S_F$ of each device, lies within 0.83--1.20 times the nominal value (5th--95th percentile).

\subsection{Comparison and Limitations}
Table~\ref{tab:compare} compares the design with reported micronewton sensors. Compared with the polymer-cladded microring array on a rigid substrate \cite{Safarloo2026}, the released membrane raises the sensitivity by more than an order of magnitude, at the cost of one sensing point per membrane. EMI immunity is expected from the all-optical readout, and wavelength-domain multiplexing on one bus is possible in principle; neither has been demonstrated, and the 0.96-nm scatter of the resonance separation restricts the pre-allocation of wavelength channels. The model errors of $S_F$ are systematic and are stated as bounds rather than standard uncertainties: the semi-vectorial solver offset (3.2\%) and the plate--continuum deviation (up to 2.5\%) give at most 5.7\% when added linearly, besides the known $0.88\%$ bus-induced sensitivity reduction. The random fabrication spread (10.3\%, $1\sigma$) is separate; both are removed by the one-point calibration of each device. All performance figures are numerical predictions. Edge compliance, the etch profile, stress relaxation, and the absorbed loss fraction are not modeled; experimental validation can build on released-membrane microring processes \cite{Leinders2015}.

\section{Conclusion}
\label{sec:conclusion}
A differential dual-microring force sensor on twin released SiO$_2$ membranes was designed using a multiphysics model that includes the anisotropic photoelastic response of silicon, the ring stiffener, and the compressive BOX prestress. The model was verified by two independent FE models, and the bus waveguide reduces the sensitivity by approximately 0.9\%. A prestress-aware stability rule selects a 5-\si{\micro\meter}-radius ring on a 17.5-\si{\micro\meter}-radius membrane, giving a bus-inclusive sensitivity of $-1.50$~pm/\si{\micro\newton}, nonlinearity below 0.05\% over 0--50~\si{\micro\newton}, a fitting-limited detection limit of 0.27~\si{\micro\newton} at 30~dB SNR (approximately 0.34~\si{\micro\newton} at system level under the assumed operating conditions), and a residual thermal cross-sensitivity of $-0.13$~\si{\micro\newton}/K. None of 1000 Monte Carlo devices buckled or showed resonance overlap, and 93.7\% met the buckling-factor margin of 3. The photoelastic effect reduces the response of compressed silicon rings, and, under the adopted stability and laser-span constraints, Vernier magnification does not exceed the sensitivity of the differential configuration. Future work will focus on refining the mechanical model to account for membrane-edge compliance and BOX-stress relaxation.


\end{document}